\documentclass[aps,prl,twocolumn,amssymb,groupedaddress]{revtex4}
\usepackage{graphicx}

\begin{document}


\title{Maximum Angle of Stability of a Wet Granular Pile}
\author{Sarah Nowak,$^\dagger$ Azadeh Samadani,$^{\dagger, *}$ and Arshad Kudrolli$^{*}$}
\affiliation{$^\dagger$Department of Physics, Massachusetts Institute of Technology, Cambridge, Massachusetts 02139\\
$^*$Department of Physics, Clark University, Worcester, Massachusetts 01610}
\date{\today}

\begin{abstract}

\end{abstract}



\maketitle

{\bf 

Anyone who has built a sandcastle recognizes that the addition of liquid to granular materials increases their stability.  However, measurements of this increased stability often conflict with theory and with each other~\cite{Hornbaker1997,Albert1997,Halsey1998,Fraysse1999,Nase2001,Samadani2001,Tegzes2002}.  A friction-based Mohr-Coulomb model has been developed~\cite{Halsey1998,Nedderman}.  However, it distinguishes between granular friction and inter-particle friction, and uses the former without providing a physical mechanism.   Albert, {\em et al.}~\cite{Albert1997} analyzed the geometric stability of grains on a pile's surface.  The frictionless model for dry particles is in excellent agreement with experiment.  But, their model for wet grains overestimates stability and predicts no dependence on system size. Using the frictionless model and performing stability analysis within the pile, we reproduce the dependence of the stability angle on system size, particle size, and surface tension observed in our experiments.  Additionally, we account for past discrepancies in experimental reports by showing that sidewalls can significantly increase the stability of granular material.   

}

The experimental apparatus consists of a clear plexiglass drum which is rotated about an horizontal axis. We primarily use a drum with a diameter $D$ of 28.5\,cm and a width $W$ which can be varied from 0\,cm to 14.5\,cm. A drum with $D = 12.5$\,cm and $W = 11.5$\,cm is also used to vary system size. The rotation rate $\omega$ is varied from $9.0 \times 10^{-4}$\,rpm to $5.6 \times 10^{-1}$\,rpm. Soda-lime glass spheres which have a density $\rho$ of 2.4\,g\,cm$^{-3}$, and with radius $r =$ 0.25\,mm, 0.3\,mm, 0.5\,mm, and 1.5\,mm and size dispersity within 0.1\,mm were used. The drum was 40\% filled with grains premixed with a small amount of liquid. We report the amount of liquid added in terms of the volume fraction, $V_f$, which is defined as the volume of the liquid divided by the volume occupied by the grains alone. The effect of surface tension of the liquid $\Gamma$ was tested by using silicone oil and water which have $\Gamma= 20 \pm 1$ dyne\,cm$^{-1}$ and $\Gamma= 70 \pm 1$ dyne\,cm$^{-1}$, respectively. Silicone oil with viscosity $\nu$ ranging from 5\,cS to 1000\,cS is used to study its impact on the measured inclination angles.  

The drum is back lit so that light could pass through a thin layer of grains that tends to accumulate on the sides of the drum, but is unable to pass through the bulk of the pile. Images were acquired with a mega-pixel resolution digital camera at a rate of three frames per second and used to determine the surface slope with automated code.  At this rate, the slope of the pile changed no more than 0.2 degrees between frames.  This error in measurement is not significant given that the slope of the pile at the moment of avalanche was distributed over about two degrees in a given run. At least 30 events were recorded to measure the mean values. 

\begin{figure}
\includegraphics[width=0.75\linewidth]{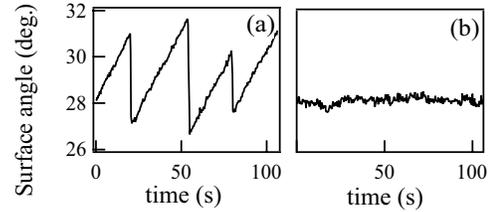}
\caption{{\bf Avalanching regimes} Glass spheres (radius $r$ 0.5 mm) 
mixed with silicone oil in a rotating drum (width $W$ 14.5 cm). (a) 
Stick-slip flow is observed for a rotation rate $\omega$ of 0.028 rpm and 
viscosity $\nu =5$ cS. (b) Continuous flow is observed for rotation rate 
of 0.28 rpm.}
\label{fig1}
\end{figure}

A stick-slip avalanche regime or a continuous avalanche regime is observed depending on the rotation rate of the drum and the liquid viscosity [see Fig.~\ref{fig1}(a,b)].  In the stick-slip regime, the slope of the pile increases linearly with time until the heap reaches the maximum angle $\theta_m$.  At this point, the grains are observed to avalanche and the pile's angle decreases to the angle of repose, $\theta_r$ when avalanching stops. In the continuous avalanche regime, the grains were observed to flow continuously and the slope of the pile is approximately constant over time. By changing $\omega$ and $\nu$, one can go from the stick-slip to continuous avalanche regime. $\theta_m$ in the stick-slip regime is observed to be independent of $\omega$ and $\nu$ within the fluctuations in the data and thus rate independent. On the other hand, $\theta_r$ and the surface angle in the continuous avalanche do depend on the viscosity, consistent with previous observations~\cite{Samadani2001}. The overall behavior as a function $\omega$ is similar to previous systematic investigations with a rotated drum~\cite{Tegzes2002}, but the magnitudes are significantly lower. 

\begin{figure}
\includegraphics[width=.7\linewidth]{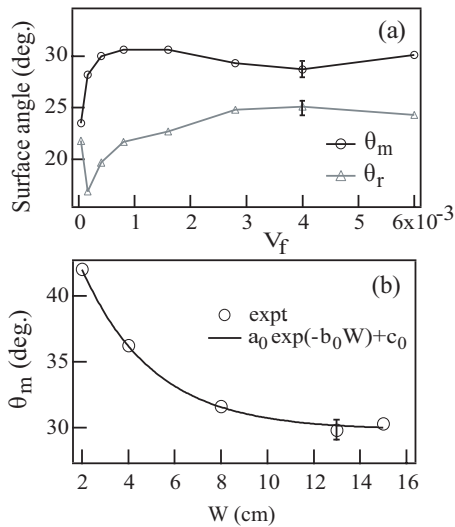}
\caption{{\bf Impact of liquid volume fraction and drum width on stability} (a) Measured surface angles as a function of $V_f$ of silicone oil. ($\nu = 5$ cS, $r = 0.5$\,mm, $\omega = 0.028$ rpm.) (b) $\theta_m$ is observed to decay exponentially as the width of the drum is increased. Error bar is indicated on one of the points. ($V_f=8 \times 10^{-3}$, $a_0 = 23.3$\,deg., $b_0 = 0.33$\,cm$^{-1}$, $c_0 = 30$\,deg.)}
\label{fig2}
\end{figure}

We note that empirically, as function of $V_f$, $\theta_m$ increases sharply, and saturates when $V_f$ is on the order $1\times 10^3$, $\theta_m$ [See figure~\ref{fig2}(a)].  The dependence of the angles on $V_f$ is again qualitatively consistent with previous studies~\cite{Tegzes2002,Samadani2001}, but as noted earlier the overall increase and saturation is significantly lower. At very small $V_f$, $\theta_m$ can be quite sensitive to $V_f$ because of the inherent roughness of the grains and its impact on the shape of the liquid bridge~\cite{Hornbaker1997}. We performed our experiments in the saturation regime, so that such effects are eliminated, but we also chose a small enough $V_f$ that the liquid does not drain to the bottom due to gravity, which would cause spatial inhomogeneity.

We experimentally determined that higher angles can be attributed to significant wall effects when narrow widths are used as in most previous investigations. For example, the angles reported in Ref.~\cite{Tegzes2002} are 15-20 degrees higher than those we report for similar $r$ and $\Gamma$, but the apparatus was 3.2 $\rm{cm}$ wide.  The observed dependence on $W$ is plotted in Fig.~\ref{fig2}(b) and is described by an exponential fit. The effect of side walls on grains which are fully immersed in a liquid has been reported in Ref.~\cite{Courrech2003}. The data in that case reaches the asymptotic value for a $W$ of a few grain diameters because liquid bridges are absent. However, the side walls have an effect over many hundred times the diameter of the particle in the partially saturated case because of the added particle-particle correlations induced by the liquid bridges. In order to simplify the analysis, we examine the data only in the limit $W > 11.5$\,cm where the side walls are unimportant.  

$\theta_m$ of at least 90 degrees is predicted by Albert, {\em et al.}~\cite{Albert1997} if the ratio of the capillary to gravitational force, which is also called the bond number ({\em Bo}~\cite{Nase2001}), is of order one or greater. Now the capillary force due to the liquid bridge bond is given by: 
\begin{equation}
F_c = \alpha \pi \Gamma r
\label{f_cap}
\end{equation}
where, $\alpha$ is a dimensionless constant which depends on the size and shape of the liquid bridge between the particles, and their separation~\cite{Lian1993,Groger2003}. Although there is some ambiguity on the value of $\alpha$, it is of order 1, and therefore {\em Bo} is approximately 5 for glass beads with $r = 0.5$\,mm mixed with silicone oil at $V_f = 0.008$. We do not observe $\theta_m$ reach 90 degree; the piles typically avalanche when $\theta$ exceeds $30$ degrees as can be noted from Fig.~\ref{fig1}. We believe that the discrepancy arises from the assumption that a wet granular pile fails at the surface, as was also noted previously in Ref.~\cite{Halsey1998}.  

\begin{figure}
\includegraphics[width=0.9\linewidth]{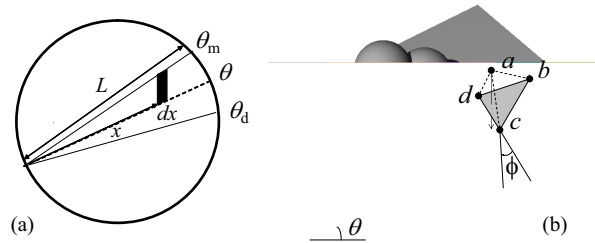}
\caption{{\bf Schematic of the liquid bridge model} (a) A schematic of the system, note that $L$ is defined as the length of the pile's surface. (b) a geometrical arrangement of a sphere resting on top of three spheres in contact used in the stability analysis. Liquid bridges located at the points of contact introduce forces along the segments joining the vertices {\em abcd} of the tetrahedron at which the spheres are centered.}
\label{fig3}
\end{figure}

Following Albert, {\em et al.}~\cite{Albert1997}, let us consider the stability of a sphere placed on top of three base spheres which are all in contact with each other (see Fig.~\ref{fig3}). In the absence of liquid bridges, the sphere becomes unstable when gravitation force is outside the triangle formed by the base spheres. By averaging over all possible orientation angle $\phi$ of the base triangle between $0$ and $60$ degree, they showed that the average inclination $\theta_d$ (the geometric, or dry angle of stability) is 23.8 degrees. At this angle, approximately half of the triangular bases at the surface can support a sphere and thus the pile will be stable. Measurements with dry glass, plastic or steel spheres in a rotated drum all show $\theta_m$ which are within a few degrees of their prediction, and thus it appears that the complications associated with random packing and friction between particles need not be taken into account to obtain reasonable results. 

Next, let us again consider a sphere resting on three base spheres as in Fig.~\ref{fig3}(b) which is located at a distance $x$ and angle $\theta$ from the bottom of the inclined surface. For $\theta > \theta_d$, the sphere will be gravitationally unstable. Now we assume that on an average, it is the component of the liquid bridge bond {\em ab} directed up the inclined plane tilted at $\theta$ which is responsible for offsetting the unbalanced gravitation component down the plane. (We neglect the contributions of the bonds {\em ac} and {\em ad} because they are mostly directed towards the axis around which the top sphere rotates when it becomes unstable.)

\begin{figure}
\includegraphics[width=0.65\linewidth]{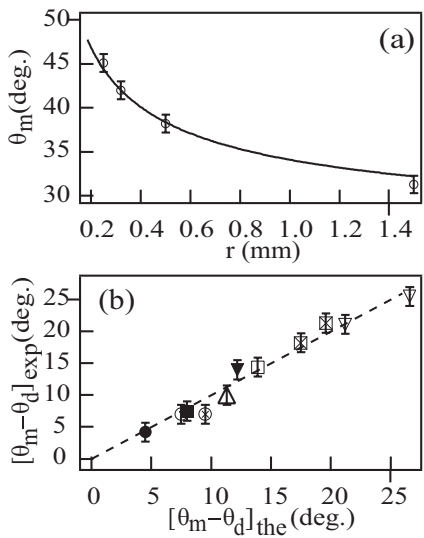}
\caption{\label{fig4} {\bf Comparison of data with liquid bridge model} (a) The maximum angle of stability glass spheres mixed with water. ($L = 27$\,cm). The solid line shows the fit to eq.~(\ref{theequation}) with $\alpha_0 = 1.2$. (b) The experimentally measured $\theta-\theta_d$ are plotted against the theoretical values calculated from eq.~(\ref{theequation}).  Key: $\circ$ - silicone oil in the large drum, $\square$ - water in the large drum, $\triangle$ - silicone oil in the small drum, $\triangledown$ - water in the small drum.  Filled shapes, open shapes, and shapes with a $\times$ correspond to $r = 1.5$\,mm, $0.5$\,mm, and $0.3$\,mm, respectively.}
\end{figure}

Because {\em abcd} forms a tetrahedron, the projection of the force $F_c$ up the plane of the base triangle due to the liquid bridge between {\em ab} can be easily determined. The average component of the liquid bond force corresponding to the average orientation angle $\phi$ of the base triangle is given by $F_c / \sqrt{24} \tan \theta_d$ which is approximately $0.46 F_c$. 
Therefore the shear stress that can be supported by the liquid bridges along the $\theta$ plane is given by the force per sphere calculated times the average number of spheres per unit area. Now, the number of particles per unit volume is $3 f_p / 4 \pi r^3$, where $f_p$ is the packing fraction of the grains which is approximately 0.64 for spherical grains. Therefore the number of particles per unit area is equal to the number of particles per unit volume to the $\frac{2}{3}$ power.  We then have the shear stress as:  
\begin{equation} \label{eq:f/a}
\frac{F_c}{\sqrt{24} \tan \theta_d} \left(\frac{3 f_p}{4 \pi r^3}\right)^{2/3},
\end{equation}
where, $F_c$ is given by eq.~(\ref{f_cap}). 

Now let us calculate the shear stress due to the unbalanced weight of a thin vertical volume element above a plane tilted at $\theta$ and located at a distance $x$ from the bottom of the surface as shown in Fig.~\ref{fig3}(a). We can write the unbalanced weight to be  
given by:
\begin{equation}
 M g \sin{(\theta-\theta_d)},
\label{w_eq}
\end{equation} 
where, $M$ is the mass of the volume element which has a length $dx$, width $W$, and a height $x \sin{(\theta_m - \theta)}/\cos{\theta_d}$.  We take $(\theta-\theta_d)$ in eq.~(\ref{w_eq}) because the pile is geometrically stable up to $\theta_d$. Thus we find that the shear stress on the $\theta$ plane from the weight of the volume element is 
\begin{equation}
x \rho g f_p \sin{(\theta_m-\theta)} \sin{(\theta-\theta_d)}/\cos{\theta_d} .
\label{noapprox}
\end{equation}

To simplify further discussion, we take the small angle approximation, and obtain the unbalanced stress to be
\begin{equation}
x \rho g f_p (\theta_m-\theta) (\theta-\theta_d)/\cos{\theta_d}.
\label{wstress}
\end{equation}
Now for each $x$ it can be shown that this shear stress is maximum for $\theta = \frac{\theta_m + \theta_d}{2}$. This is simply because as one goes above $\theta_d$ on one hand there is less materials above, but on the other hand the volume element is tilted further off equilibrium. 

Next, it is not clear if the pile will fail when the shear stress from the weight first exceeds the stress provided by the liquid bridges, or when the average shear stress due to the weight of the pile exceeds the stress provided by the liquid bridges. In reality, the correct physical situation is most likely combination of these two. Therefore, we set $x = \beta L$ in eq.~(\ref{wstress}) when balancing the stresses, where $\beta$ is a dimensionless constant between 0.5 and 1. $\beta = 1$, if adjacent volume elements cannot offset any stress by a particular volume element, and $\beta = 0.5$, if the stress is evenly distributed along the slip plane. Given the variance seen in the stick-slip events [see Fig.~\ref{fig1}(a)], and the heterogeneity of pile at the granular level, $\beta$ may vary from event to event. 

Combining these facts and balancing the two stress components, we obtain the condition for equilibrium as
\begin{equation}
\left(\frac{\theta_m-\theta_d}{2}\right)^2=\left(\frac{9\pi}{16 f_p}\right)^{\frac{1}{3}}\left(\frac{\alpha \cos{\theta_d} \Gamma}{\sqrt{24} \tan{\theta_d} \rho g r \beta L}\right)\,,
\end{equation}
and, setting $\alpha_0 = \alpha/\beta$, we obtain
\begin{equation}
\theta_m - \theta_d = 
\sqrt{
\left(\frac{9\pi}{2 f_p}\right)^{\frac{1}{3}}\left(\frac{\alpha_0 \cos{\theta_d} \Gamma}{\sqrt{6} \tan{\theta_d} \rho g r L}\right)}\,\,\,.
\label{theequation}
\end{equation}

Figure~\ref{fig4}(a) shows a comparison of the measured $\theta_m$ as a function of particle size $r$ and the fit to eq.~(\ref{theequation}) after conversion of units to degrees. The data is well described by using $\alpha_0 = 1.2 \pm 0.05$. To further test the model, we measured $\theta_m$ in two different sized systems ($L = 27$\,cm and $L = 12.5$\,cm) with various combinations of $\Gamma$ and $r$. Fig.~\ref{fig4}(b) shows a plot of the experimentally measured $\theta_m$ above $\theta_d$ compared with eq.~(\ref{theequation}), with $\alpha_0 = 1.2$. Excellent overall agreement is observed.  

We note that the small angle approximation used to simplify eq.~(\ref{noapprox}) is reasonable over the angles tested. The analysis can be extended to higher angles without using this approximation, but $\theta_m$ has to be obtained by a numerical solution of the corresponding equation. In developing our analysis, we have ignored the contribution of friction which will also help stabilize the wet pile, but just as in the dry case, we may expect the friction contribution to be small as a grain rolls as it gets dislodged from the triangular base. A more detailed knowledge of how the failure develops may clarify this issue.

In conclusion, our analysis captures the observed stability dependence on grain size, system size, and surface tension.  It would be of great interest to consider how well this analysis extends to piles composed of non-spherical and multi-sized grains as in natural sand.  


\bibliographystyle{nature}
\bibliography{sn-bib}

Author to whom correspondence and requests for
materials should be sent: A. Kudrolli.

\begin{acknowledgments}
{\bf Acknowledgments} We thank J. Norton and N. Israeloff for their help with the apparatus, and J. Bico for feedback on the manuscript. The work was supported by the National Science Foundation Grant No. DMR-9983659, and the GLUE program of the Department of Energy. 
\end{acknowledgments}

\end{document}